\documentclass[11pt,english,a4paper,floatfix]{article}
\usepackage{jcappub}
\usepackage{multirow}
\usepackage[normalem]{ulem}
\pdfoutput=1
\usepackage{bm}
\usepackage{amsfonts}
\usepackage[utf8]{inputenc}
\usepackage{lscape}  
\usepackage{booktabs}  
\usepackage{rotating} 
\usepackage{pdflscape} 
\usepackage{orcidlink}
\graphicspath{{Figures/}}

\usepackage{graphicx}
\usepackage{amssymb}
\usepackage{amsmath}
\usepackage{times} 
\usepackage{multirow}
\usepackage{url}
\usepackage{cuted}
\usepackage[english]{babel}
\usepackage{enumerate}
\usepackage[normalem]{ulem}
\usepackage{enumitem}
\usepackage{multirow}
\usepackage{float}
\usepackage{booktabs}
\usepackage{makecell}
\usepackage{placeins}
\usepackage{newtxtext,newtxmath}
\usepackage[T1]{fontenc}
\usepackage{amsmath}	
\usepackage{csquotes}

\begin{document}
	
	
	\title{Detection prospects of solar $g$-modes with LISA}

    \author[a]{Aman Awasthi \orcidlink{0000-0003-3568-6471}}

	\affiliation[a]{Department of Physics, Indian Institute of Technology Bombay, Powai, Mumbai, Maharashtra 400076, India}
	\emailAdd{174120011@iitb.ac.in}


\abstract{
 The possibility of detecting solar oscillation modes using space-based gravitational-wave detectors has been investigated in the context of gravitational-wave interferometry, with Polnarev \cite{Polnarev:2009xf} demonstrating that low-frequency solar modes could, in principle, produce detectable signals in a LISA-type interferometer. Motivated by this work, I revisit the problem using current solar models and updated detector sensitivities, and in addition to observational constraints, I consider theoretical predictions for mode amplitudes. In this study, I compute the gravitational response of solar oscillation modes using standard solar models generated with \texttt{MESA}, and mode eigenfrequencies and eigenfunctions calculated with \texttt{GYRE}. I focus primarily on solar $g$ modes, evaluating their responses for degree $l=2$ and azimuthal orders $m=0$ and $m=2$. The analysis incorporates both the earlier proposed and the current updated LISA sensitivity curves, and I perform a comparative assessment with the Taiji and TianQin missions in the relevant low-frequency band. To assess the robustness of the predicted signals, I estimate the gravitational responses using two different standard solar models based on the GS98 and AGSS09 abundance compilations. I find that the resulting signal responses are nearly identical for the two models, indicating that uncertainties in solar metallicity have a negligible impact on the detectability of solar $g$ modes by space-based interferometers.
}

\maketitle



\section{Introduction}

The Sun, as a dynamic and highly complex star, exhibits a wide range of oscillations and waves \cite{Christensen-Dalsgaard_2002RvMP...74.1073C}, \cite{Deubner_1984ARA&A..22..593D}, \cite{Christensen_Dalsgaard_n_Berthomieu_1991sia..book..401C} that provide invaluable insights into its internal structure and dynamics. These oscillations are broadly categorized into different types of modes based on the dominant restoring force which controls the oscillations. Among these, pressure ($p$) modes, where the pressure gradient provides the restoring force, and gravity ($g$) modes, where the buoyancy provides the restoring force, are two of the most important classes of solar oscillation modes.  These modes are primarily studied through helioseismology, which is the study of the propagation of seismic waves within the Sun. The Sun's structure, composition, and energy transport mechanisms can be inferred by analyzing these oscillations.

Unlike p-modes, which are primarily sensitive to the solar structure outside the inner core, g-modes probe the radiative region below the outer convection zone. $g$-modes in the Sun are characterized by much lower frequencies ($< 500\;\mu$Hz). The study of $g$-modes is of particular importance for understanding the Sun's internal rotation and the structure of its energy generating core. These modes are characterized by three quantum numbers, the radial order $n$, the degree $l$, and azimuthal order $m$.

Extensive efforts over the last 50 years using the velocity or intensity at the solar surface have given a wide spectrum of p-modes \cite{Claverie_1979_Natur.282..591C}, \cite{Duvall_and_Harvey_1983Natur.302...24D}, \cite{Libbrecht_1990ApJS...74.1129L}, whereas a similar success is not achieved in the $g$ modes detection. Early efforts in helioseismology identified gravity ($g$) modes as promising probes of the solar core, leading to several early detection claims based on ground-based observations \cite{Severnyi_1976Natur.259...87S},\cite{Brookes_1976Natur.259...92B}. However, none of these claims were confirmed by subsequent long-term observations, including more than a decade of data from the Solar and Heliospheric Observatory (SOHO), which instead provided only upper limits on $g$-mode amplitudes \cite{Appourchaux_2000ApJ...538..401A,Appourchaux_2010}. 

Later analyses of low-frequency helioseismic data reported signatures consistent with high-order $g$ modes \cite{Garcia_2007Sci...316.1591G}, but these results were also not independently confirmed \cite{Appourchaux_2010}. More recently, indirect evidence for solar $g$ modes was claimed based on temporal variations in the large frequency separation of acoustic modes \cite{Fossat_2017}. This interpretation, however, has been challenged by independent reanalyses and theoretical studies, which questioned both the robustness of the signal extraction and the physical interpretation \cite{Schunker_2018SoPh..293...95S,Appourchaux_Corbard_2019}. Consequently, solar $g$ modes remain observationally unconfirmed, motivating alternative detection strategies.

Cutler and Lindblom \cite{Cutler:1996ai} were the first to explore the external gravitational perturbations associated with solar modes on LISA. Later, Polnarev et al\cite{Polnarev:2009xf},  investigated the response of a spaceborne laser interferometer, such as the Laser Interferometer Space Antenna (LISA), to solar oscillations. In that pioneering work, it was shown that low-frequency solar oscillations could induce measurable time-dependent gravitational perturbations in a space-based detector. Two distinct contributions were identified: a near-zone (Newtonian) contribution arising from the time-varying gravitational quadrupole moment of the Sun, and a far-zone contribution associated with gravitational radiation emitted by the oscillating mass distribution. Polnarev\cite{Polnarev:2009xf} demonstrated that, in the low-frequency regime relevant for solar oscillations, the near-zone contribution dominates and that, under favorable conditions, a LISA-type detector could achieve a higher signal-to-noise ratio than traditional helioseismic measurements.

Following the publication of Polnarev’s work, the LISA \cite{amaro2017laser},\cite{Danzmann2011} mission concept has matured substantially, with updated and more realistic sensitivity curves that incorporate instrumental noise as well as astrophysical backgrounds, such as the Galactic binary confusion noise. In addition, new space-based interferometer concepts, such as Taiji \cite{TaijiScientific:2021qgx}, \cite{Ruan_2020_Taiji_sensitivity}, and  TianQin \cite{TianQin_Luo_2016}, have been proposed, providing complementary sensitivity in overlapping frequency bands. These developments motivate a re-examination of the detectability of solar oscillation modes using space-based gravitational wave detectors, incorporating updated solar models, updated detector sensitivities. In this work, I revisit the problem originally addressed by Polnarev\cite{Polnarev:2009xf}, with particular emphasis on solar $g$ modes, and extend the analysis in several important directions.

Given the ongoing debate about the solar composition, I explicitly examine the impact of the solar abundance by considering two different standard solar models based on the GS98 \cite{Grevesse&Sauval_1998SSRv...85..161G} and AGSS09 \cite{Asplund_2009} heavy-element abundance compilations. The difference between these solar models should cover the expected uncertainty due to solar models for g-mode detection. 

I also evaluate the expected gravitational wave signal strengths under two distinct assumptions regarding the mode amplitudes. I consider conservative upper bounds inferred from helioseismic velocity measurements by the GOLF \cite{Gabriel_1997SoPh..175..207G} experiment, as well as theoretically predicted amplitudes derived from excitation models\cite{Balmforth_1992MNRAS.255..639B}. This dual approach allows us to assess both optimistic and pessimistic detection scenarios and to clarify the conditions under which space-based detection may be feasible.

Finally, I compute the detector response using the current updated LISA sensitivity curve, including both instrumental noise and the Galactic binary confusion background, and I compare the results with the projected sensitivity of the Taiji \cite{Ruan_2020_Taiji_sensitivity} TianQin \cite{TianQin_Hu_2018} mission.

In addition to estimating total signal responses, I separately analyze the near-zone and far-zone contributions to the gravitational signal. This allows us to identify the dominant physical mechanisms across the relevant frequency range and to clarify the role of gravitational radiation relative to Newtonian perturbations for solar oscillation modes.

The structure of this paper is as follows. In Section \ref{sec:solar_model}, I describe the construction of the standard solar models using \texttt{MESA} and the computation of oscillation modes with \texttt{GYRE}. My results for the gravitational signal responses for various models are presented in section \ref{sec:detector}. In section \ref{sec:near_and_far_zone} near and far zone signals are approximated, and signal-to-noise ratios are discussed in section \ref{sec:SNR_LISA_estimation}. Finally, I summarize my findings and discuss their implications for helioseismology and future space-based observations in Section ~\ref{sec:conclusion}.


\section{Standard solar models using MESA and the computation of oscillation modes} \label{sec:solar_model}
A standard solar model (SSM) is constructed by evolving a $1M_\odot$ star through the known solar age and matching the luminosity, radius, and surface $Z/X$ to the observed values. This is achieved by adjusting the initial $Y$, $Z$, and the mixing length parameter. The constraints on solar luminosity and radius are particularly important. As a result, the SSM has no free parameters, but different SSMs can nevertheless be obtained by adopting different input physics, such as opacities and nuclear reaction rates, as well as different chemical compositions.

The stellar evolution calculations presented in this work were carried out using MESA (Modules for Experiments in Stellar Astrophysics), a widely used open-source code for one-dimensional stellar evolution developed by Bill Paxton and collaborators \cite{Polnarev:2009xf},\cite{Paxton_2011ApJS..192....3P},\cite{Paxton_2013ApJS..208....4P},\cite{Paxton_2015ApJS..220...15P},\cite{Paxton_2018ApJS..234...34P},\cite{Paxton_2019ApJS..243...10P}. MESA solves the fully coupled equations governing stellar structure and evolution while incorporating modern input physics such as nuclear reaction networks, radiative opacities, and equations of state.

In this study, I computed two standard solar models, Updated MESA GS98 and Updated MESA AGSS09, and investigated their mode sensitivity in LISA. The main updates in these models are the change of four nuclear reaction rates relevant for hydrogen burning in the stellar core.  These updates are motivated from the recent development in the solar models studies \cite{Vinyoles_2017}

The reaction rates are updated by changing the $S(0)$ factor in the reaction rate formula in \texttt{MESA}.  The quantity $S(0)$ represents the zero-energy astrophysical S-factor, which parameterizes the intrinsic nuclear cross section after removing the strong energy dependence caused by Coulomb barrier penetration. Modifying $S(0)$ therefore directly alters the normalization of the nuclear reaction rate used in the stellar evolution calculations.

These reactions and their updated values of the $S(0)$  are as follows: 
   
    \begin{itemize}
        \item For $ \rm{p(p, e^{+} \nu_{e})d}$, chosen S factor is  $ \rm{S_{11}(0) = (4.03 \pm 0.006)\times 10^{-25} MeV b} $ \cite{Marcucci_2013}
        \vspace{0.1cm}
        \item  For $ \rm{^7Be (p, \gamma)^8 B}$, chosen S factor is  $ \rm{S_{17}(0) = (2.13 \pm 0.01)\times 10^{-5} MeV b} $  \cite{ZHANG2015535}, \cite{Adelberger_2011}
        \vspace{0.1cm}
        \item   For $ \rm{^{14}N (p, \gamma)^{15} O}$, chosen S factor is  $ \rm{S_{114}(0) = 1.59\times 10^{-3} MeV b} $ \cite{Marta_2011PhRvC..83d5804M}
        \vspace{0.1cm}
         \item For $ \rm{^{3}He (^4 He, \gamma)^{7} Be}$, chosen S factor is $ \rm{S_{34}(0) = (5.72 \pm 0.12)\times 10^{-4} MeV b} $ \cite{Iliadis_2016} 
    \end{itemize}
Apart from these selected reactions rate changes, the equation of state used in the stellar evolution is Free\_EOS \cite{Alan_Irwin_FreeEOS_2012ascl.soft11002I}. In this study, I construct two solar models, both of them are obtained from the \texttt{simplex\_solar\_calibration} model in the mesa-24.08.1\cite{Paxton_2011ApJS..192....3P}  \texttt{test\_suite}. The specific assumptions in each of these models are listed as below
\begin{itemize}
    \item \textbf{Updated MESA GS98}: 
    The values of input parameters for this model is mentioned in the Table \ref{tab:solar_models_input_output}.  Here, surface $ (Z/X ) = 0.02292$  corresponds to the solar abundance of  Grevesse $\&$ Sauval (GS98)\cite{Grevesse&Sauval_1998SSRv...85..161G}  \\
    
    \item \textbf{Updated MESA AGSS09}:  The values of input parameters for this model is mentioned in the Table \ref{tab:solar_models_input_output}. Here, surface $ (Z/X ) = 0.0181$ corresponds to the solar abundance of  AGSS09 \cite{Asplund_2009}.  
    
\end{itemize}

\begin{figure}[] 
    \centering
    \includegraphics[width=0.8\textwidth]{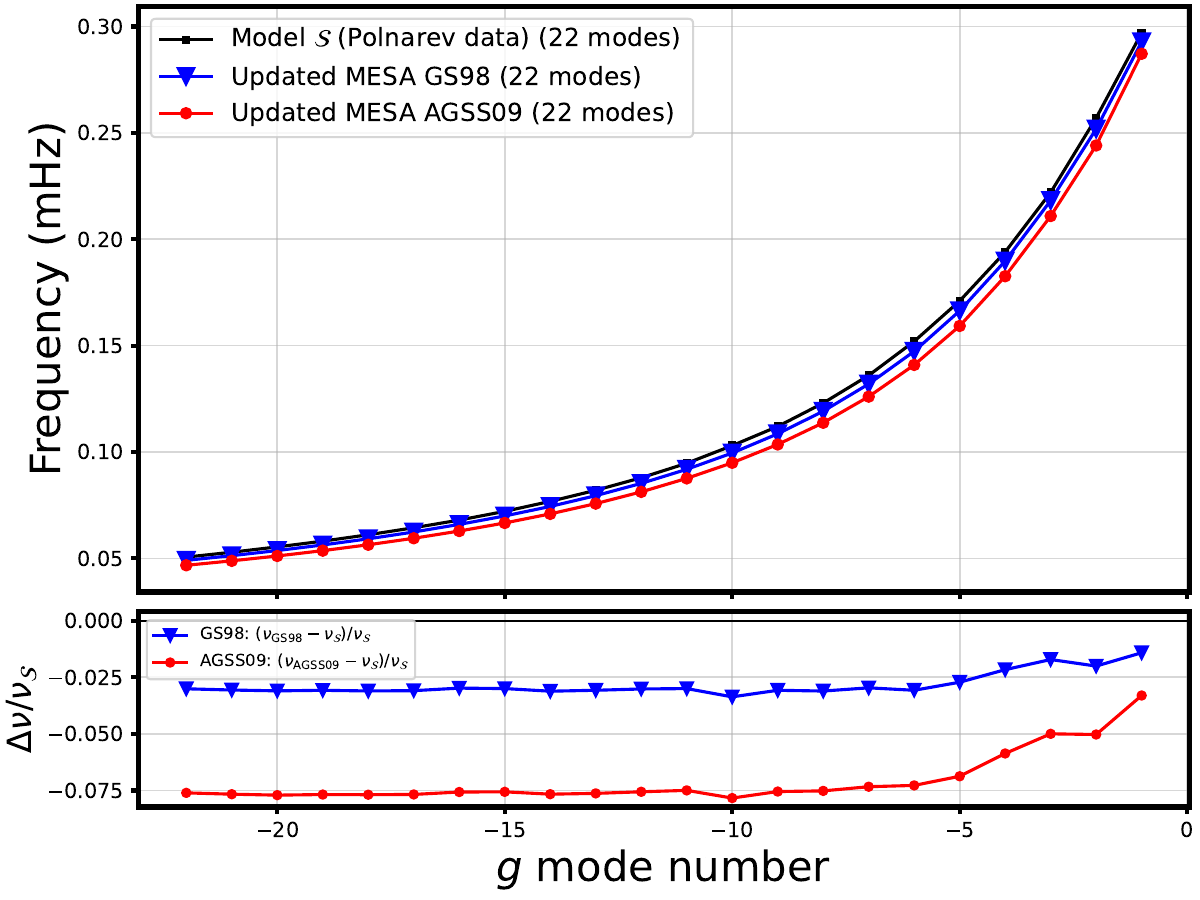} 
    \caption{Comparison of the $g$ mode frequencies from the Updated  MESA GS98, Updated MESA AGSS09 models and \cite{Polnarev:2009xf} values. The first 22 solar $g$ modes, along with the fractional differences relative to the models are shown.} 
    \label{fig:freq_compare_upd_MESA_2_ZbyX_n_Pol} 
\end{figure}

\begin{figure}[] 
    \centering
    \includegraphics[width=0.8\textwidth]{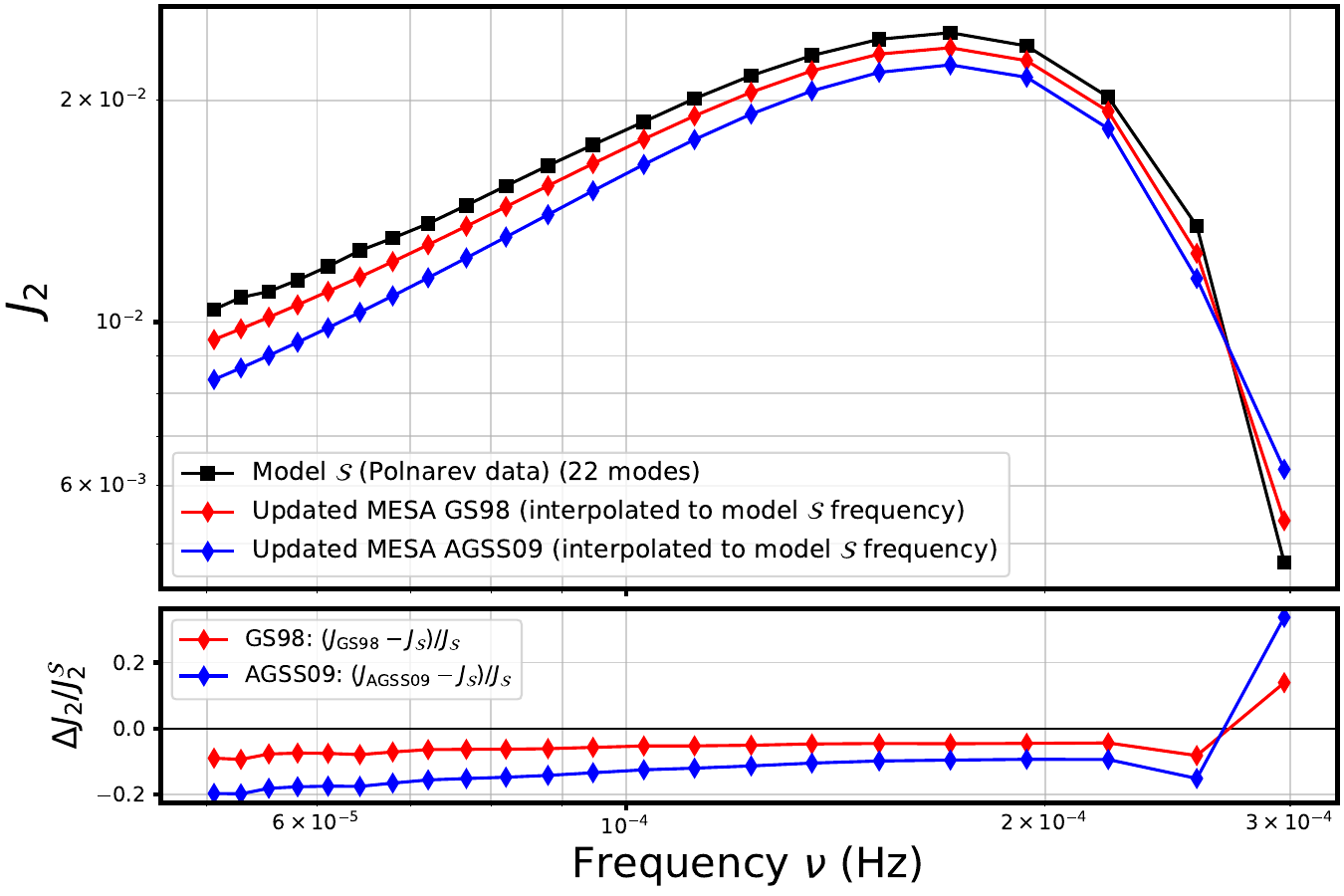} 
    \caption{Comparison of the quadrupole moment $J_2$ values for solar $g$  modes from the Updated  MESA GS98, Updated MESA AGSS09 models and \cite{Polnarev:2009xf} values. The first 22 solar $g$ modes, along with the fractional differences relative to the models are shown.} 
    \label{fig:Jnm_comparison_upd_MESA_2_ZbyX_n_Pol} 
\end{figure}

\begin{table}[htbp]
\centering
\renewcommand{\arraystretch}{1.3} 
\caption{Input parameters and observable properties for four standard solar models computed with MESA.}
\label{tab:solar_models_input_output}
\begin{tabular}{|p{3.9cm}|p{1.0cm}|p{1.0cm}|p{1.0cm}|p{1.9cm}|p{1.7cm}|p{1.4cm}|}
\hline
\multirow{2}{*}{\textbf{Model}} & \multicolumn{3}{c|}{\textbf{Input Parameters}} & \multicolumn{3}{c|}{\textbf{Observables}} \\
\cline{2-7}
 & $Y_{\rm{ini}}$  & $Z_{\rm{ini}}$ & $\alpha$ & $\rm{log}\left( L/L_{\odot}\right)$ & $\rm{log}\left( R/R_{\odot}\right)$ & $  (Z/X)_{\rm{surf}}$ \\
\hline
Updated MESA GS98 & 0.2664 & 0.0183 & 1.8852  & $2.142 \times 10^{-7} $ & $6.36\times 10^{-7} $  & 0.0229  \\
\hline
Updated MESA AGSS09 &  0.2472  & 0.0149  & 1.7886 & $ 4.792 \times 10^{-7}$ &$7.72\times 10^{-7} $ & 0.0181   \\
\hline
\end{tabular}
\end{table}
I use these models in GYRE\cite{Townsend_2013MNRAS.435.3406T}  to calculate the eigen frequencies and eigen functions for the solar oscillation modes. GYRE solves the boundary value problem of stellar oscillations using advanced numerical techniques, enabling accurate determination of mode frequencies and spatial eigenfunctions across a broad frequency range. The frequency and quadrupole moments obtained from these models for the first 22 solar $g $ modes is shown in the Figure (\ref{fig:freq_compare_upd_MESA_2_ZbyX_n_Pol}) and Figure  (\ref{fig:Jnm_comparison_upd_MESA_2_ZbyX_n_Pol}), respectively. I compare these results  with the previously estimated values\cite{Polnarev:2009xf}, which used the model $\mathcal{S}$\cite{CSD_1996Sci...272.1286C}. 

The comparative analysis of the frequencies from the two models is shown in Figure~(\ref{fig:freq_compare_upd_MESA_2_ZbyX_n_Pol}). The figure presents the $g$-mode frequencies as a function of $g$-mode number. The upper panel shows the frequencies of the three models (\textit{Updated MESA AGSS09}, \textit{Updated MESA GS98}, and Model $\mathcal{S}$) versus the mode number, while the lower panel displays the fractional differences of frequencies of the \textit{Updated MESA AGSS09} and \textit{Updated MESA GS98} models relative to Model $\mathcal{S}$.
Both the \textit{Updated MESA GS98} and \textit{Updated MESA AGSS09} models reproduce the overall trend of the reference frequencies of the model $\mathcal{S}$ well. However, small systematic differences are present. The \textit{Updated MESA GS98} model shows fractional deviations in the range of approximately $1$–$3\%$, while the \textit{Updated MESA AGSS09} model exhibits fractional deviations in the range of approximately $3$–$8\%$. The fractional deviations for both models become nearly constant at lower $g$-mode frequencies.

The comparative analysis of the quadrupole moments of the modes of the two models is shown in Figure~(\ref{fig:Jnm_comparison_upd_MESA_2_ZbyX_n_Pol}). 
The upper panel shows the quadrupole moments of solar $g$ modes of the three models (\textit{Updated MESA AGSS09}, \textit{Updated MESA GS98}, and Model $\mathcal{S}$) versus frequency, while the lower panel displays the fractional differences of quadrupole moments of the \textit{Updated MESA AGSS09} and \textit{Updated MESA GS98} models relative to Model $\mathcal{S}$. Both models reproduce the overall trend of the reference quadrupole moments of the modes for model $\mathcal{S}$ well. However, small systematic differences are observed. The \textit{Updated MESA GS98} model shows fractional deviations in the range of approximately $10$–$20\%$, while the \textit{Updated MESA AGSS09} model exhibits deviations in the range of about $20$–$40\%$. The fractional deviations in the quadrupole moments for both models become nearly constant at lower $g$-mode frequencies.

I  provide in Table \ref{tab:modes_data} a summary of the results obtained for the values of relevant quantities derived from the solar oscillation data for the Updated MESA GS98 model. Although the properties of more than 50 modes were analyzed, only the first 25 modes are presented, as they are representative and capture the essential behavior observed across all modes. The columns list the following quantities: radial order $n$; mode frequency $\nu$; horizontal displacement eigenfunction at the solar surface $\xi_h$;  quadrupole moment $J_2$; total kinetic energy of the mode $E$; gravitational signals for azimuthal number $ m = 0, \pm2$; velocity amplitudes for azimuthal number  $m = 0, \pm 2$; and the expected instrumental noise, binary confusion noise surface velocity amplitudes respectively, for the LISA mission at the corresponding frequencies (details are in Section \ref{sec:detector}).  

All quantities reported here are scaled to an assumed mode amplitude and therefore do not represent absolute physical values. Such normalization is required because, in linear oscillation theory, the eigenfunctions are defined only up to an arbitrary multiplicative constant. In the final calculations, the velocity amplitudes are normalized to match the observational limits or the theoretically predicted values.

\begin{table}[ht]
\centering
\resizebox{\textwidth}{!}{%
\begin{tabular}{rlllllllllll}
\toprule
$n $& $\nu$ & $ \xi_h$ & $J_2$ & $E$ & $S_0$ & $S_2$ & $V_0$ & $V_2$ & $B_{I}$ & $B_{b}$ & $B_{v}$ \\
\midrule
\midrule
-1 & $2.93\times 10^{-4}$ & $1.17\times 10^{-1}$ & $5.38\times 10^{-3}$ & $2.04\times 10^{40}$ & $6.16\times 10^{-16}$ & $3.38\times 10^{-15}$ & $1.60\times 10^{5}$ & $1.96\times 10^{5}$ & $2.25\times 10^{-18}$ & $2.20\times 10^{-18}$ & $1.45\times 10^{1}$ \\
-2 & $2.52\times 10^{-4}$ & $1.57\times 10^{-1}$ & $1.41\times 10^{-2}$ & $3.26\times 10^{40}$ & $1.51\times 10^{-15}$ & $1.18\times 10^{-14}$ & $1.49\times 10^{5}$ & $1.82\times 10^{5}$ & $3.36\times 10^{-18}$ & $2.53\times 10^{-18}$ & $1.49\times 10^{1}$ \\
-3 & $2.18\times 10^{-4}$ & $2.10\times 10^{-1}$ & $2.02\times 10^{-2}$ & $5.10\times 10^{40}$ & $2.21\times 10^{-15}$ & $2.24\times 10^{-14}$ & $1.41\times 10^{5}$ & $1.73\times 10^{5}$ & $4.96\times 10^{-18}$ & $3.10\times 10^{-18}$ & $1.67\times 10^{1}$ \\
-4 & $1.90\times 10^{-4}$ & $2.77\times 10^{-1}$ & $2.31\times 10^{-2}$ & $9.99\times 10^{40}$ & $2.78\times 10^{-15}$ & $3.37\times 10^{-14}$ & $1.37\times 10^{5}$ & $1.67\times 10^{5}$ & $7.29\times 10^{-18}$ & $3.59\times 10^{-18}$ & $1.67\times 10^{1}$ \\
-5 & $1.66\times 10^{-4}$ & $3.61\times 10^{-1}$ & $2.37\times 10^{-2}$ & $1.75\times 10^{41}$ & $3.29\times 10^{-15}$ & $4.48\times 10^{-14}$ & $1.35\times 10^{5}$ & $1.65\times 10^{5}$ & $1.06\times 10^{-17}$ & $4.35\times 10^{-18}$ & $1.77\times 10^{1}$ \\
-6 & $1.47\times 10^{-4}$ & $4.60\times 10^{-1}$ & $2.29\times 10^{-2}$ & $2.77\times 10^{41}$ & $3.78\times 10^{-15}$ & $5.52\times 10^{-14}$ & $1.35\times 10^{5}$ & $1.65\times 10^{5}$ & $1.49\times 10^{-17}$ & $4.96\times 10^{-18}$ & $1.93\times 10^{1}$ \\
-7 & $1.32\times 10^{-4}$ & $5.74\times 10^{-1}$ & $2.16\times 10^{-2}$ & $4.10\times 10^{41}$ & $4.24\times 10^{-15}$ & $6.48\times 10^{-14}$ & $1.37\times 10^{5}$ & $1.68\times 10^{5}$ & $2.05\times 10^{-17}$ & $5.64\times 10^{-18}$ & $1.97\times 10^{1}$ \\
-8 & $1.19\times 10^{-4}$ & $7.03\times 10^{-1}$ & $2.00\times 10^{-2}$ & $5.84\times 10^{41}$ & $4.71\times 10^{-15}$ & $7.38\times 10^{-14}$ & $1.40\times 10^{5}$ & $1.72\times 10^{5}$ & $2.75\times 10^{-17}$ & $6.34\times 10^{-18}$ & $2.07\times 10^{1}$ \\
-9 & $1.09\times 10^{-4}$ & $8.48\times 10^{-1}$ & $1.86\times 10^{-2}$ & $8.01\times 10^{41}$ & $5.16\times 10^{-15}$ & $8.23\times 10^{-14}$ & $1.45\times 10^{5}$ & $1.77\times 10^{5}$ & $3.62\times 10^{-17}$ & $7.15\times 10^{-18}$ & $2.04\times 10^{1}$ \\
-10 & $9.95\times 10^{-5}$ & $1.01$ & $1.72\times 10^{-2}$ & $1.07\times 10^{42}$ & $5.62\times 10^{-15}$ & $9.06\times 10^{-14}$ & $1.50\times 10^{5}$ & $1.84\times 10^{5}$ & $4.66\times 10^{-17}$ & $8.03\times 10^{-18}$ & $2.13\times 10^{1}$ \\
-11 & $9.19\times 10^{-5}$ & $1.18$ & $1.60\times 10^{-2}$ & $1.39\times 10^{42}$ & $6.08\times 10^{-15}$ & $9.88\times 10^{-14}$ & $1.56\times 10^{5}$ & $1.91\times 10^{5}$ & $5.90\times 10^{-17}$ & $8.80\times 10^{-18}$ & $2.18\times 10^{1}$ \\
-12 & $8.52\times 10^{-5}$ & $1.37$ & $1.49\times 10^{-2}$ & $1.81\times 10^{42}$ & $6.55\times 10^{-15}$ & $1.07\times 10^{-13}$ & $1.62\times 10^{5}$ & $1.99\times 10^{5}$ & $7.35\times 10^{-17}$ & $9.52\times 10^{-18}$ & $2.22\times 10^{1}$ \\
-13 & $7.95\times 10^{-5}$ & $1.58$ & $1.39\times 10^{-2}$ & $2.28\times 10^{42}$ & $7.03\times 10^{-15}$ & $1.15\times 10^{-13}$ & $1.69\times 10^{5}$ & $2.07\times 10^{5}$ & $9.04\times 10^{-17}$ & $9.95\times 10^{-18}$ & $2.25\times 10^{1}$ \\
-14 & $7.44\times 10^{-5}$ & $1.80$ & $1.31\times 10^{-2}$ & $2.85\times 10^{42}$ & $7.52\times 10^{-15}$ & $1.24\times 10^{-13}$ & $1.76\times 10^{5}$ & $2.16\times 10^{5}$ & $1.10\times 10^{-16}$ & $1.11\times 10^{-17}$ & $2.41\times 10^{1}$ \\
-15 & $6.99\times 10^{-5}$ & $2.04$ & $1.24\times 10^{-2}$ & $3.51\times 10^{42}$ & $8.03\times 10^{-15}$ & $1.32\times 10^{-13}$ & $1.83\times 10^{5}$ & $2.25\times 10^{5}$ & $1.32\times 10^{-16}$ & $1.19\times 10^{-17}$ & $2.46\times 10^{1}$ \\
-16 & $6.60\times 10^{-5}$ & $2.29$ & $1.18\times 10^{-2}$ & $4.30\times 10^{42}$ & $8.56\times 10^{-15}$ & $1.41\times 10^{-13}$ & $1.91\times 10^{5}$ & $2.34\times 10^{5}$ & $1.57\times 10^{-16}$ & $1.26\times 10^{-17}$ & $2.41\times 10^{1}$ \\
-17 & $6.24\times 10^{-5}$ & $2.56$ & $1.12\times 10^{-2}$ & $5.23\times 10^{42}$ & $9.10\times 10^{-15}$ & $1.50\times 10^{-13}$ & $1.99\times 10^{5}$ & $2.44\times 10^{5}$ & $1.85\times 10^{-16}$ & $1.38\times 10^{-17}$ & $2.68\times 10^{1}$ \\
-18 & $5.92\times 10^{-5}$ & $2.85$ & $1.07\times 10^{-2}$ & $6.29\times 10^{42}$ & $9.67\times 10^{-15}$ & $1.60\times 10^{-13}$ & $2.07\times 10^{5}$ & $2.53\times 10^{5}$ & $2.17\times 10^{-16}$ & $1.43\times 10^{-17}$ & $2.65\times 10^{1}$ \\
-19 & $5.63\times 10^{-5}$ & $3.15$ & $1.03\times 10^{-2}$ & $7.47\times 10^{42}$ & $1.02\times 10^{-14}$ & $1.69\times 10^{-13}$ & $2.15\times 10^{5}$ & $2.63\times 10^{5}$ & $2.51\times 10^{-16}$ & $1.54\times 10^{-17}$ & $2.71\times 10^{1}$ \\
-20 & $5.37\times 10^{-5}$ & $3.47$ & $9.90\times 10^{-3}$ & $8.86\times 10^{42}$ & $1.09\times 10^{-14}$ & $1.79\times 10^{-13}$ & $2.23\times 10^{5}$ & $2.73\times 10^{5}$ & $2.90\times 10^{-16}$ & $1.58\times 10^{-17}$ & $2.84\times 10^{1}$ \\
-21 & $5.13\times 10^{-5}$ & $3.80$ & $9.56\times 10^{-3}$ & $1.04\times 10^{43}$ & $1.15\times 10^{-14}$ & $1.90\times 10^{-13}$ & $2.32\times 10^{5}$ & $2.84\times 10^{5}$ & $3.32\times 10^{-16}$ & $1.79\times 10^{-17}$ & $2.87\times 10^{1}$ \\
-22 & $4.91\times 10^{-5}$ & $4.15$ & $9.25\times 10^{-3}$ & $1.23\times 10^{43}$ & $1.21\times 10^{-14}$ & $2.01\times 10^{-13}$ & $2.40\times 10^{5}$ & $2.94\times 10^{5}$ & $3.79\times 10^{-16}$ & $1.78\times 10^{-17}$ & $2.89\times 10^{1}$ \\
-23 & $4.71\times 10^{-5}$ & $4.51$ & $8.98\times 10^{-3}$ & $1.42\times 10^{43}$ & $1.28\times 10^{-14}$ & $2.12\times 10^{-13}$ & $2.49\times 10^{5}$ & $3.05\times 10^{5}$ & $4.29\times 10^{-16}$ & $1.91\times 10^{-17}$ & $2.95\times 10^{1}$ \\
-24 & $4.52\times 10^{-5}$ & $4.89$ & $8.73\times 10^{-3}$ & $1.65\times 10^{43}$ & $1.35\times 10^{-14}$ & $2.23\times 10^{-13}$ & $2.57\times 10^{5}$ & $3.15\times 10^{5}$ & $4.84\times 10^{-16}$ & $1.98\times 10^{-17}$ & $2.98\times 10^{1}$ \\
-25 & $4.35\times 10^{-5}$ & $5.28$ & $8.51\times 10^{-3}$ & $1.91\times 10^{43}$ & $1.42\times 10^{-14}$ & $2.35\times 10^{-13}$ & $2.66\times 10^{5}$ & $3.26\times 10^{5}$ & $5.44\times 10^{-16}$ & $2.05\times 10^{-17}$ & $3.03\times 10^{1}$ \\
\bottomrule
\end{tabular}}
\caption{The frequency $\nu$ (Hz) and the respective amplitudes of horizontal surface displacement $\xi_{h}(R_{\odot})$, quadrupole moment $J$, the total kinetic energy of the mode $E$ (kg.m$^2$/sec$^2$), gravitational signal $S_0$, $S_2$ and velocity amplitudes $V_0$ (m/sec) and $V_2$ (m/sec), for unit radial surface displacement $\xi_r(R_{\odot} = 1)$ as a function of the radial order of the mode $n$, (negative sign corresponds to $g$ modes). The columns $B_I$ $(1/\sqrt{\rm{Hz}})$,  $B_b$ $(1/\sqrt{\rm{Hz}})$  and $B_v$ $(\rm{m/sec}/\sqrt{\rm{Hz}})$ are the expected instrumental noise,  binary confusion noise surface velocity amplitudes respectively, for the LISA mission at the corresponding frequencies.}
\label{tab:modes_data}
\end{table}

\section{ Sensitivity of solar $g$ modes for space based interferometers} \label{sec:detector}

LISA \cite{amaro2017laser}, \cite{Danzmann2011}  consists of three arms, which are arranged in a triangular configuration and orbit the Sun in a circular path. This configuration is designed to measure gravitational waves by detecting tiny changes in the length of the arms due to the passage of a gravitational wave. The response of the interferometer is determined by comparing the phase shifts of the laser signals that travel along the different arms of the detector. Specifically, the response is measured by examining the difference in the fractional changes in the phase of the round-trip laser signals along two of the arms. For example, we might compare the phase shift in the signal that travels from point A to B and back (denoted as the arm ABA) with the phase shift in the signal that travels from point A to C and back (denoted as the arm ACA) \cite{Polnarev:2009xf}. 
\begin{equation}
    S = \bigg( \frac{\delta \varphi }{\varphi} \bigg)_{AB} -  \bigg( \frac{\delta \varphi }{\varphi} \bigg)_{AC} 
\end{equation}

The response $S_m$ of a space-based interferometric detector to a solar oscillation mode of azimuthal order $m$ and frequency $\nu=\omega/2\pi$ can be written, following ~\cite{Polnarev:2009xf}, as
\begin{equation}
S_m(\nu,\phi) = \frac{1}{2}\,C_m\,J_m
\left( \frac{R_\odot}{r}\right)^5
\left( \frac{\nu_\odot}{\nu}\right)^2
\left[
f^N_m
+\frac{2}{3}\left(\frac{\nu}{\nu_r}\right)^4
f^{GW}_m
\right],
\label{eq: Polnarev_Sm}
\end{equation}
where $\phi$ denotes the angular position of the detector with respect to the Sun. The  $C_m$ is the geometric coupling coefficient for modes, and $J_m$ is the quadrupole moment of the mode. The quantity $R_\odot$ is the solar radius and $r$ is the Sun--detector distance, taken to be one astronomical unit,
$r = 1.496 \times 10^{11}\,\mathrm{m}$. 

The characteristic frequency
\begin{equation}
\nu_\odot \approx \frac{1}{2\pi}\sqrt{\frac{G M_\odot}{R_\odot^3}} \approx 10^{-4}\,\mathrm{Hz},
\end{equation}
is the solar dynamical frequency, while
\begin{equation}
\nu_r \approx \frac{c}{2\pi r} \approx 3 \times 10^{-4}\,\mathrm{Hz}
\end{equation}
defines the transition frequency separating the near-zone and far-zone regimes. Here $G$ is the gravitational constant, $M_\odot$ is the solar mass, and $c$ is the speed of light.

The total detector response consists of two physically distinct contributions. The first term, $f^N_m$, represents the near-zone (Newtonian) signal arising from the time-dependent gravitational quadrupole moment of the Sun associated with the oscillation mode. In this regime, the detector is located well within one gravitational wavelength of the source, and the signal is dominated by instantaneous Newtonian tidal perturbations rather than propagating gravitational waves. The near-zone contribution is given by
\begin{equation}
f^N_m = {\cal I}_m^{\alpha\beta}\,\Delta T_{\alpha\beta},
\end{equation}
where ${\cal I}_m^{\alpha\beta}$ is the mass quadrupole moment tensor of the mode, and $\Delta T_{\alpha\beta}$ encodes the detector’s tensor response to Newtonian tidal fields.

The second term, $f^{GW}_m$, corresponds to the far-zone contribution due to gravitational radiation emitted by the oscillating quadrupole mass distribution. This contribution becomes relevant when the detector lies in the radiation zone of the source and is suppressed at low frequencies by the factor $(\nu/\nu_r)^4$. It is given by
\begin{equation}
f^{GW}_m = \tilde{\cal I}_m^{\alpha\beta}\,\Delta N_{\alpha\beta},
\end{equation}
where $\tilde{\cal I}_m^{\alpha\beta}$ denotes the transverse--traceless (TT) projection of the quadrupole tensor, and $\Delta N_{\alpha\beta}$ represents the detector response to incoming gravitational waves.

The response signal for each azimuthal mode on LISA can be expressed as \cite{Polnarev:2009xf}  
\begin{equation}
\begin{aligned}
 S_0(\nu,\phi) &= {\frac{3\sqrt{3}}{16}} C_{0}J_{0}  \frac{R_{\odot}}{r^5}   \frac{\nu_{\odot}^2 }{{\nu}^2} \bigg( 1 + \frac{7}{3} \frac{\nu^4}{\nu_r ^4} \bigg) \rm{sin}\bigg(2 \psi  + \frac{\pi}{3}\bigg)   \\
 S_2(\nu,\phi) &= 2 \sqrt{3}C_{2}J_{2}  \frac{R_{\odot}}{r^5}   \frac{\nu_{\odot}^2 }{{\nu}^2} \Bigg[ \rm{sin}2 \chi \rm{cos}\bigg(2 \psi  + \frac{\pi}{3}\bigg)    - \frac{25}{32}\bigg( 1 + \frac{7}{75} \frac{\nu^4}{\nu_r ^4} \bigg)  \rm{cos}2 \chi \rm{sin}\bigg(2 \psi  + \frac{\pi}{3}\bigg)    \Bigg]
\end{aligned}
\end{equation}
where, $\psi = - \phi$ and $ \chi \approx  13.65\phi $  and $C_0 = - \sqrt{\frac{5}{4 \pi}}$ and $  C_2  =  \sqrt{\frac{15}{4 \pi}}$ .
Explicitly, $ \chi$  is the angle between the radius vector to the detector and the instantaneous orientation of the reference axes corotating with the Sun, and  $ \psi$ is the angle between the arm AB and the direction of the orbit. 

I restrict my analysis to the $ m=0 $ and $m = \pm 2$ components (denoted $S_0$ and $S_2$), as the $m= \pm 1$ modes do not contribute to the integrated velocity signal for an observer located in the solar equatorial plane. This follows from the antisymmetry of their angular dependence about the equator, which leads to an exact cancellation of their contributions over the visible disk. Consequently, these modes are not detectable in disk-integrated Doppler measurements and are therefore excluded from the present analysis.

In the calculations presented here, I have assumed that the coupling coefficients satisfy $J_m=J_2$. For this analysis, using the updated MESA GS98 and updated MESA AGSS09 solar models, I compute and display the scaled signal overlaid on the LISA sensitivity curve.
To enable a direct comparison with the instrumental sensitivity, the signal amplitude is scaled by the frequency resolution according to
\( S_m / \sqrt{\Delta f} \),
where the frequency resolution is defined as
\[
\Delta f = \frac{1}{T_{\mathrm{obs}}},
\]
with \( T_{\mathrm{obs}} = 10 \) years expressed in seconds \cite{amaro2017laser}, \cite{ESA_LISA_MissionSummary}. This choice corresponds to the assumption that the LISA mission will operate in space for ten years.
The resulting scaled signal, obtained using this frequency resolution, is shown in the corresponding figure alongside the LISA sensitivity curve.

In addition to LISA, I also consider the proposed space-based gravitational wave detectors Taiji\cite{TaijiScientific:2021qgx},\cite{Ruan_2020_Taiji_sensitivity} Program and TianQin\cite{TianQin_Luo_2016},\cite{TianQin_Hu_2018}, both of which are designed to probe the low-frequency gravitational wave band. Taiji is a heliocentric mission concept similar in configuration to LISA, while TianQin is planned as a geocentric detector optimized for millihertz gravitational waves. Because the expected gravitational signals produced by solar $g$-modes lie in the low-frequency regime, these missions may also provide a complementary platform for investigating such signals. To examine this possibility, I compare the predicted response of the solar $g$-modes with the projected sensitivity curves of Taiji and TianQin. The corresponding sensitivity curves used in this analysis are shown in Figure (\ref{fig:final_S0_S2_vs_frequency}).


\subsection{Response of Solar $g$ Modes on Space-Based Detectors}

\begin{figure}[] 
    \centering
    \includegraphics[width=0.9\textwidth]{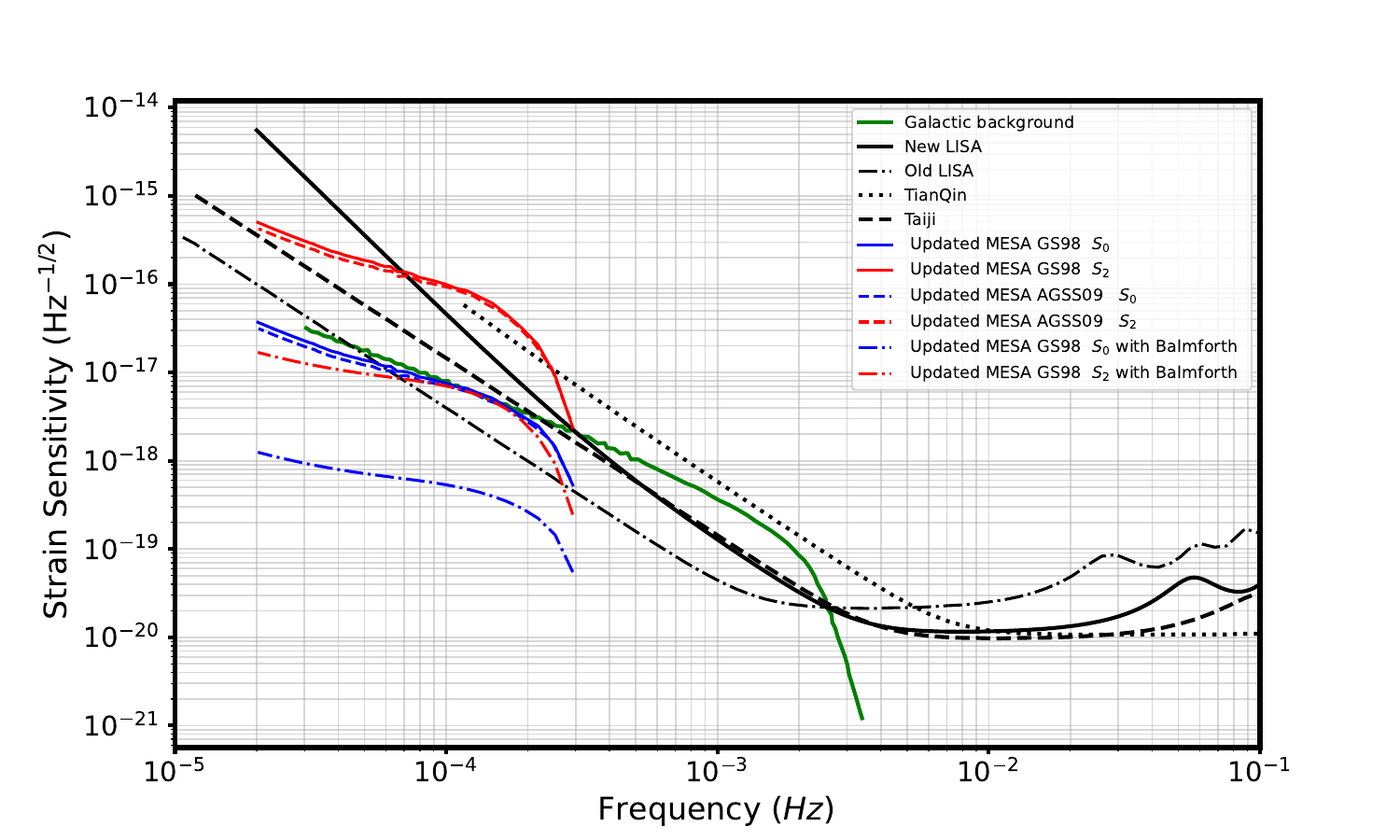} 
    \caption{Projected response of the $S_0$ and $S_2$ for a signal-to-noise ratio (SNR) of 3 in velocity-based experiments, shown for the LISA, Taiji, and TianQin sensitivities. The signals (arising from the time-varying solar quadrupole field) are computed using the Updated MESA GS98 and Updated MESA AGSS09 solar models, assuming an observational upper bound on the $g$-mode velocity amplitude. Signals corresponding to theoretically predicted velocity amplitudes are also included. For comparison, results are shown for both the originally proposed and the current updated LISA sensitivity curves.} 
    \label{fig:final_S0_S2_vs_frequency} 
\end{figure}

I investigate the response of solar oscillation modes on space-based gravitational wave detectors, with particular emphasis on the solar $g$ modes. My primary focus is on the LISA, considering both its earlier proposed sensitivity curve \cite{Larson_2000}, \cite{Bender1998LISA}, and the currently updated sensitivity \cite{amaro2017laser}. For comparison, I also include the sensitivity of Taiji\cite{Ruan_2020_Taiji_sensitivity} and TianQin \cite{TianQin_Luo_2016}, \cite{TianQin_Hu_2018} in the relevant low-frequency regime. The corresponding sensitivities and estimated signal responses are shown in Figure~(\ref{fig:final_S0_S2_vs_frequency}). In this figure, the strain sensitivity is plotted as a function of frequency. The plot includes the predicted sensitivity curves of space-based interferometers, namely the old and updated LISA configurations, as well as Taiji and TianQin. In addition, the noise contribution from the galactic foreground is shown as a continuous green curve.

I approximate the gravitational signal responses of solar $g$ modes for azimuthal orders $m = 0$ and $m = 2$, denoted by $S_0$ and $S_2$, respectively. These responses are evaluated using two different standard solar models: the \textit{Updated MESA GS98} model and the \textit{Updated MESA AGSS09} model. The use of these two models allows us to assess the impact of possible variations in the solar models on the predicted gravitational signal strengths.

The $S_0$ and $S_2$ responses for both solar models are estimated by assuming an upper bound on the $g$-mode surface velocity amplitudes, inferred from velocity measurements with signal-to-noise ratio $\mathrm{SNR} = 3$ in the GOLF \cite{Gabriel_1997SoPh..175..207G} (Global Oscillations at Low Frequencies) experiment. This choice represents a conservative but observationally motivated upper limit on the mode amplitudes.

From Figure~(\ref{fig:final_S0_S2_vs_frequency}), I observe that the $S_0$  responses obtained from the GS98 and AGSS09 models (solid blue and dashed blue curves) are nearly indistinguishable, and similarly for $S_2$ (solid red 
 and dashed red curve). This indicates that uncertainties associated with solar metallicity and abundance determinations have only a minor effect on the predicted gravitational response of solar $g$ modes. Consequently, I infer that the choice of solar model does not significantly influence the detectability of these modes by space-based interferometers.


%
A key result of the analysis is that the $S_2$ signal response for both solar models lies above the current LISA \cite{amaro2017laser} sensitivity curve within the frequency range $7 \times  10^{-5} \lesssim    \nu \lesssim 3 \times 10^{-4}\,\mathrm{Hz}$. In comparison, the $S_2$ signal response for both solar models also lies above the projected sensitivity curve of Taiji \cite{Ruan_2020_Taiji_sensitivity} for all the modes considered in this study. These results suggest that solar  $g$ modes with azimuthal order $m = 2$ may be detectable by LISA and Taiji within this frequency range, provided their amplitudes are close to the observational upper limits inferred from helioseismic velocity measurements.

In addition to these observationally constrained estimates, I also examine the gravitational signal responses derived from theoretically predicted $g$-mode velocity amplitudes, as computed by Balmforth~\cite{Balmforth_1992MNRAS.255..639B}. The corresponding signal responses are found to lie well below the current sensitivity limits of Taiji, LISA, and TianQin. This highlights the fact that detectability critically depends on whether the true solar $g$-mode amplitudes are closer to observational upper bounds rather than theoretical expectations.

Projected response of the $S_0$ and $S_2$ for a signal-to-noise ratio (SNR) of 3 in velocity-based experiments, shown for the Taiji, LISA and TianQin sensitivities. The gravitational-wave signals are computed using the Updated MESA GS98 and Updated MESA AGSS09 solar models, assuming an observational upper bound on the $g$-mode velocity amplitude. Signals corresponding to theoretically predicted velocity amplitudes are also included. For comparison, results are shown for both the originally proposed and the current updated LISA sensitivity curves.

\section{Near-Zone and Far-Zone Contributions} \label{sec:near_and_far_zone}

\begin{figure}[] 
    \centering
    \includegraphics[width=0.9\textwidth]{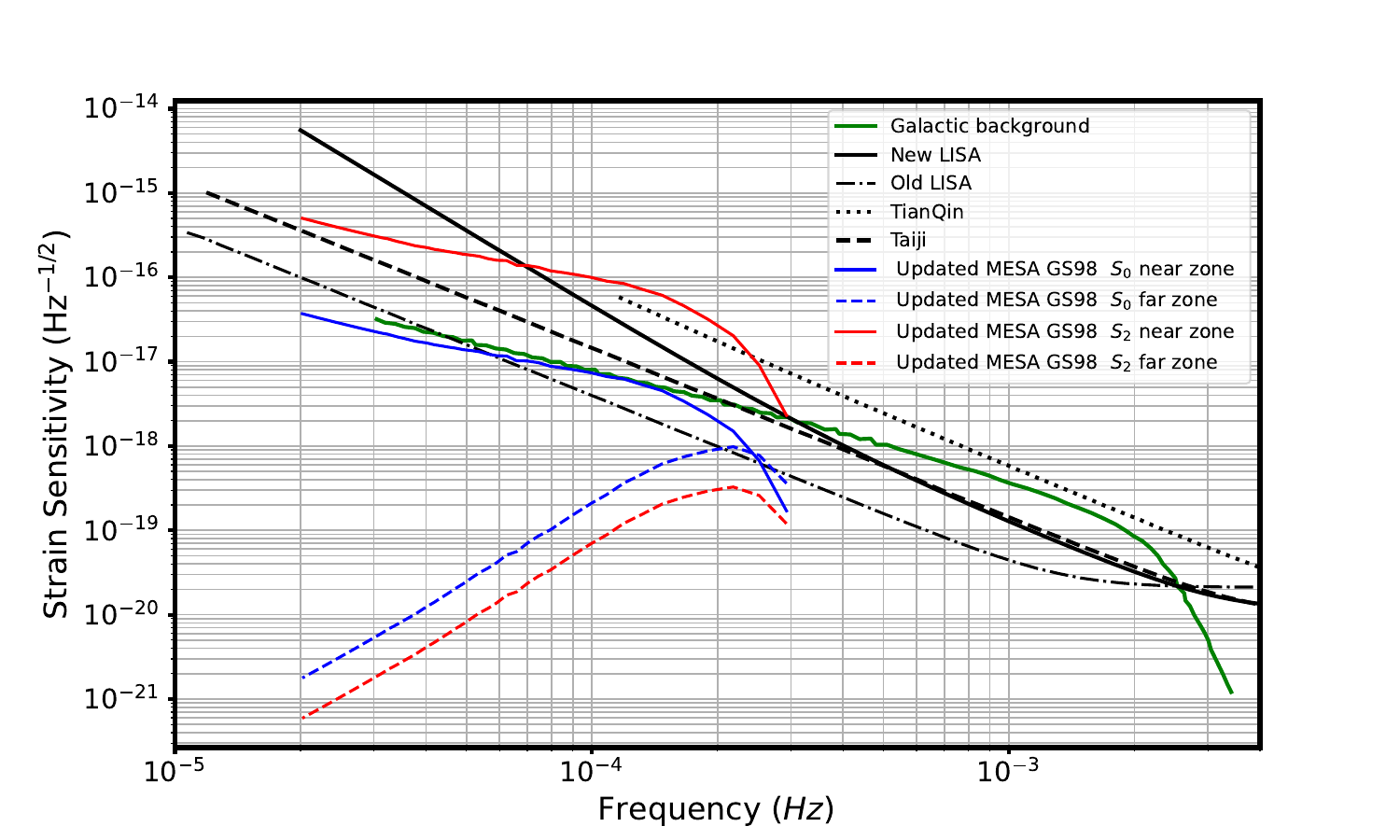} 
    \caption{Projected near-zone (solid lines) and far-zone (dashed lines) responses of the \(S_0\) and \(S_2\) modes for the Updated MESA GS98 solar model, assuming a signal-to-noise ratio (SNR) of 3 in velocity-based experiments. The responses are shown for the LISA, Taiji and TianQin sensitivities. For comparison, results corresponding to both the originally proposed and the current updated LISA sensitivity curves are included.} 
    \label{fig:near_and_far_S0_S2_vs_frequency} 
\end{figure}

The near and far zone response of the signals can be obtained from the equation \ref{eq: Polnarev_Sm} \cite{Polnarev:2009xf}. I analyze the response of solar $g$ modes on space-based detectors by separately evaluating the near-zone and far-zone contributions to the gravitational signal. The corresponding results are shown in Figure~(\ref{fig:near_and_far_S0_S2_vs_frequency}). In this figure, strain sensitivity is shown as a function of frequency. The graph presents the expected sensitivity curves for space-based interferometers, including both the original and updated configurations of LISA, along with Taiji and TianQin. Additionally, the contribution of noise from the galactic foreground is illustrated by a solid green curve. As evident from the figure, the near-zone contribution dominates over the far-zone contribution for both azimuthal orders considered. This behavior arises mainly because the far-zone contribution scales as $(\nu/\nu_r)^4$, where $\nu_r$ is a characteristic reference frequency. As a result, the far-zone term decreases rapidly with decreasing frequency, rendering it subdominant in the low-frequency regime relevant for solar oscillation modes. In contrast, the near-zone contribution does not suffer from this suppression and therefore provides the dominant contribution to the total signal.


%
In particular, the near-zone $S_2$ response lies  above the current LISA sensitivity \cite{amaro2017laser} for frequencies $ 7 \times  10^{-5}\lesssim    \nu \lesssim 3 \times 10^{-4}\,\mathrm{Hz}$. In comparison, the near-zone $S_2$ signal response also lies above the projected sensitivity curve of Taiji \cite{Ruan_2020_Taiji_sensitivity} for all the modes considered in this study. This reinforces the conclusion that solar $g$ modes with azimuthal order $m = 2$ are the most promising candidates for detection by space-based interferometers.

\section{Signal-to-Noise Ratio Estimates for LISA} \label{sec:SNR_LISA_estimation}

In this section, I estimate the SNR using the relation~\cite{Polnarev:2009xf}
\begin{equation}
\left(\frac{S} {N}\right) \approx  \frac{h_{\rm TD} \ {\sqrt{T}}}{(h_I + h_b) },
\end{equation}
where $h_{\rm TD}$ denotes the threshold detectable strain for a given signal-to-noise ratio $S/N$ in velocity experiments, $h_I$ is the instrumental noise (which is referred as $B_I$ in the Table \ref{tab:modes_data}), $h_b$ is the binary confusion noise (which is referred as $B_b$ in the Table \ref{tab:modes_data}), and $T$ is the observation time of 10 years in seconds. $h_I$ are  shown in the Figure (\ref{fig:final_S0_S2_vs_frequency})  for the detector and $h_b$ is represented by the solid green curve.

\begin{figure}[] 
    \centering
    \includegraphics[width=0.9\textwidth]{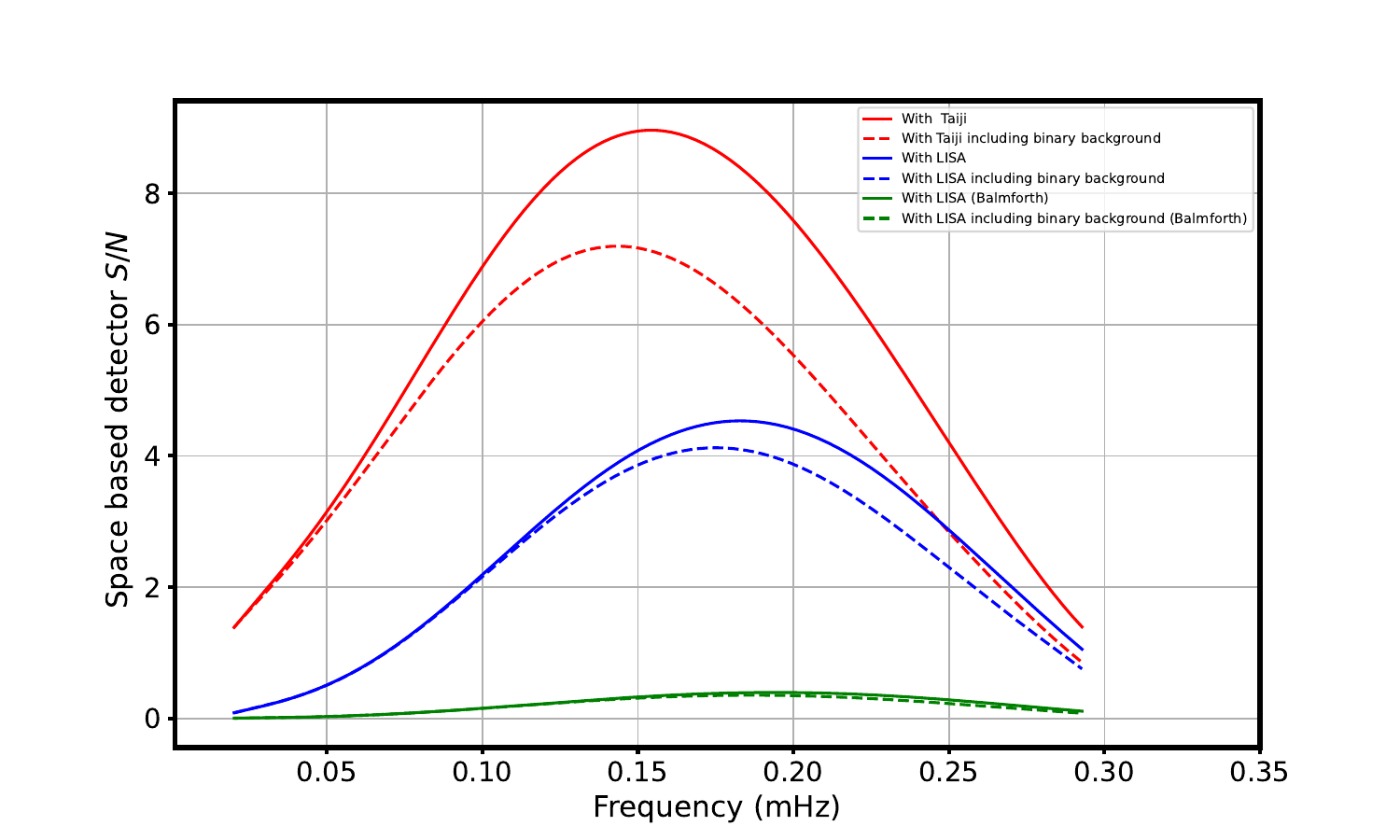} 
    \caption{Comparison of the signal-to-noise ratio ($S/N$) for LISA and Taiji as a function of frequency for solar oscillation modes with $\ell=2$ and $m=2$. The $S/N$ is evaluated assuming a detection threshold of $S/N=3$ for solar velocity measurements. For Taiji, the solid and dashed red curves correspond to instrumental noise only and to instrumental noise combined with Galactic binary confusion noise, respectively, assuming the velocity-based upper-limit amplitude. For LISA, the solid and dashed blue curves show the corresponding $S/N$ for the same upper-limit amplitude. The solid and dashed green curves represent the LISA $S/N$ computed using the theoretically predicted velocity amplitudes from Ref.~\cite{Balmforth_1992MNRAS.255..639B}, again for instrumental noise only and for instrumental plus Galactic binary confusion noise, respectively.} 
    \label{fig:LISA_SNR_with_frequency_new_comparison} 
\end{figure}

 I present a comparison of the expected signal-to-noise ratio (SNR) for LISA\cite{amaro2017laser} and Taiji\cite{Liu_2023_Taiji}, as a function of frequency, shown in Figure~(\ref{fig:LISA_SNR_with_frequency_new_comparison}), for solar oscillation modes with $\ell=2$ and $m=2$. The SNR is evaluated under two different assumptions for the solar mode amplitudes: (i) conservative upper limits inferred from helioseismic velocity measurements with $\mathrm{SNR}=3$, and (ii) theoretically predicted velocity amplitudes from excitation models~\cite{Balmforth_1992MNRAS.255..639B}. In all cases, I use the updated LISA instrumental noise model and include relevant astrophysical background contributions.

For each amplitude assumption, two noise scenarios are considered. In the first case, only instrumental noise is included, shown by the solid blue curve for the velocity-based upper-limit amplitudes and by the solid black curve for the theoretically predicted amplitudes. In the second case, both instrumental noise and the Galactic binary confusion noise are included, shown by the dashed red and dashed green curves, respectively.

This comparison illustrates that while the inclusion of confusion noise significantly degrades sensitivity at lower frequencies, the SNR corresponding to the velocity-based upper-limit amplitudes remains appreciable over a broad frequency range. In contrast, the theoretically predicted amplitudes yield SNR values well below the LISA detection threshold across the full frequency band considered.

Under the assumption that LISA and Taiji are independent detectors and hence have independent noise, but observe at the same time. I can combine the $SNR$ from them in quadrature sum\cite{Pai_2001} as 
\begin{equation}
    SNR_{Net} = \sqrt{SNR_{LISA}^2 + SNR_{Taiji}^2}.
\end{equation}

\begin{figure}[] 
    \centering
    \includegraphics[width=0.9\textwidth]{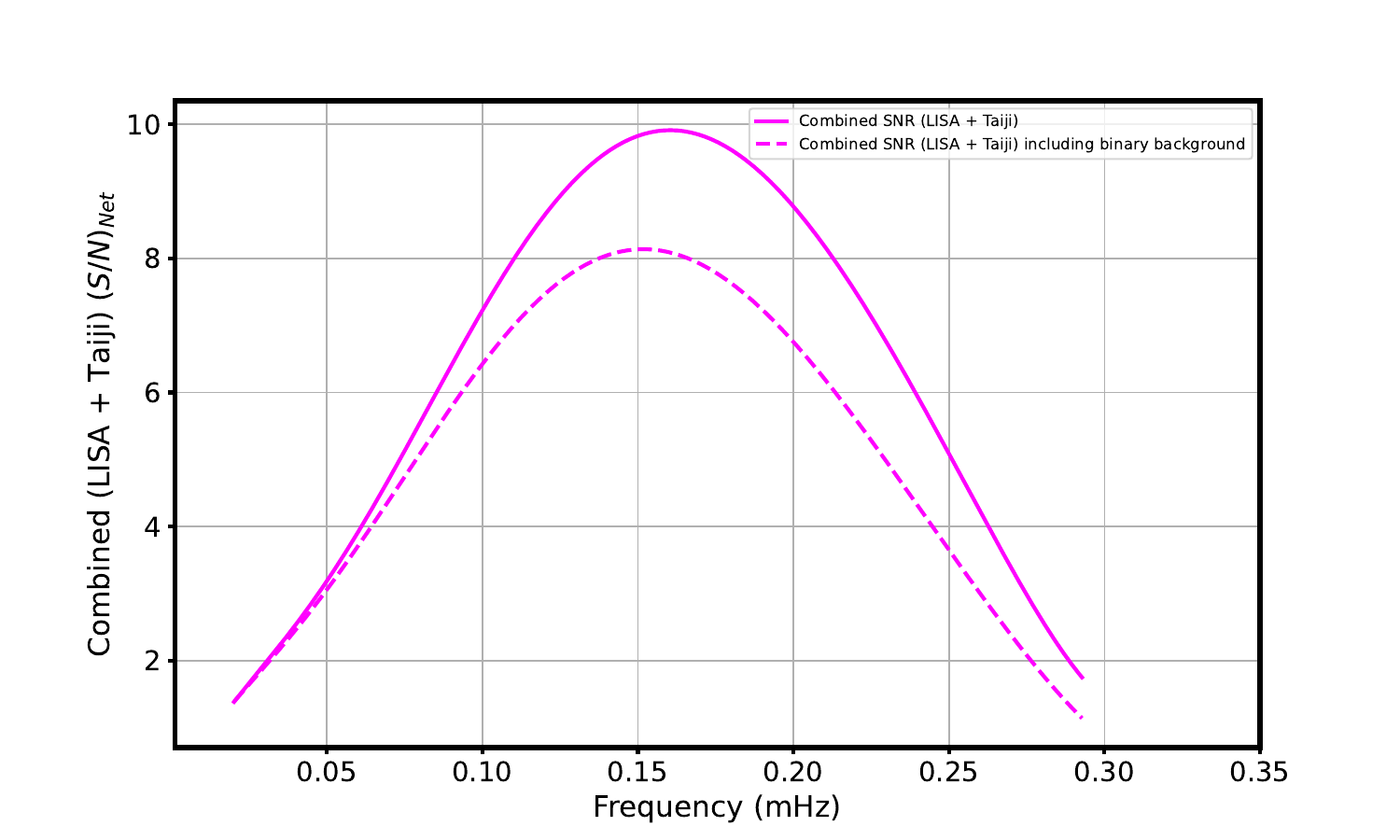} 
    \caption{Combined signal-to-noise ratio ($S/N$) for LISA and Taiji as a function of frequency for solar oscillation modes with $\ell=2$ and $m=2$. The $S/N$ is evaluated assuming a detection threshold of $S/N=3$ for solar velocity measurements. The solid-magenta and dashed-magenta curves correspond to instrumental noise only and to instrumental noise combined with Galactic binary confusion noise, respectively.  } 
    \label{fig:Combined_SNR_LISA_and_Taiji} 
\end{figure}

I also present the combined signal-to-noise ratio (SNR)  for LISA\cite{amaro2017laser} and Taiji\cite{Liu_2023_Taiji}, \cite{Ruan_2020_Taiji_sensitivity} detectors as a function of frequency for solar oscillation modes with $\ell=2$ and $m=2$.  The SNR is computed assuming a detection threshold of $S/N=3$ consistent with the upper bound on the velocity amplitude for solar velocity measurements. In the analysis, two noise scenarios are considered: instrumental noise alone and instrumental noise combined with the Galactic binary confusion noise. The corresponding results are illustrated by the solid-magenta and dashed-magenta curves in Figure (\ref{fig:Combined_SNR_LISA_and_Taiji}), respectively.

\section{Conclusion} \label{sec:conclusion}
In this work, I investigate the feasibility of detecting solar oscillation modes with space-based gravitational wave detectors, focusing in particular on solar $g$ modes and their gravitational response. The analysis is carried out using the sensitivity of LISA \cite{Larson_2000} and the revised LISA\cite{amaro2017laser} configuration. I further assess the detectability of these modes with other proposed missions, namely Taiji\cite{Ruan_2020_Taiji_sensitivity} and TianQin\cite{TianQin_Hu_2018}, and also explore the potential improvement obtained from the combined sensitivity of the revised LISA and Taiji detectors.


The gravitational responses of solar $g$ modes were evaluated for degree $l=2$ and azimuthal orders $m = 0$ $(S_0)$ and $m = 2$ $(S_2)$, using two different standard solar models, namely the \textit{Updated MESA GS98} and \textit{Updated MESA AGSS09} models. These models differ primarily in their adopted surface metal abundances during the evolutionary calculations. In addition, they incorporate updates to four nuclear reaction rates and employ different choices of the equation of state, as detailed in Section \ref{sec:solar_model}. By adopting conservative upper bounds on the surface velocity amplitudes inferred from GOLF observations with signal-to-noise ratio greater than three, I found that the predicted $S_0$ and $S_2$ responses from the two solar models are nearly identical. This indicates that uncertainties associated with the solar abundance problem have a negligible impact on the gravitational detectability of solar $g$ modes.

A key outcome of this study is that the gravitational response associated with the $S_2$ component for both solar models exceeds the current LISA sensitivity \cite{amaro2017laser} within the frequency range $7 \times 10^{-5} \lesssim \nu \lesssim 3 \times 10^{-4}\,\mathrm{Hz}$. Moreover, the $S_2$ signal response for both models remains above the projected sensitivity curve of Taiji \cite{Ruan_2020_Taiji_sensitivity} for all modes examined in this work. This suggests that these modes may be detectable by these detectors if their true amplitudes are close to the observational upper limits. In contrast, the responses derived from theoretically predicted velocity amplitudes are found to lie well below the sensitivity thresholds of updated LISA, Taiji, and TianQin, emphasizing the critical role of mode amplitudes in determining detectability.

I also examined the near-zone and far-zone contributions to the gravitational response of solar $g$ modes. This analysis shows that the near-zone contribution dominates over the far-zone contribution throughout the frequency range of interest  $2 \times 10^{-5} \lesssim \nu \lesssim 3 \times 10^{-4}\,\mathrm{Hz}$. This dominance arises from the suppression of the far-zone term at low frequencies, where it scales proportionally to $(\nu/\nu_r)^4$. Notably, the analysis shows that the near-zone $S_2$ response exceeds the present LISA sensitivity \cite{amaro2017laser} in the frequency interval $7 \times 10^{-5} \lesssim \nu \lesssim 3 \times 10^{-4}\,\mathrm{Hz}$. For Taiji \cite{Ruan_2020_Taiji_sensitivity}, the near-zone $S_2$ signal response lies above the projected sensitivity curve for all modes considered in this work. This behavior indicates that solar $g$ modes with azimuthal order $m = 2$ represent the most promising targets for detection with future space-based interferometric detectors. Such a response effectively probes the temporal variations in the Newtonian gravitational potential.

Finally, I present signal-to-noise ratio estimates for LISA and Taiji as a function of frequency, incorporating updated instrumental noise as well as the Galactic binary confusion noise. The two detectors are assumed to operate independently, and the signal-to-noise ratio of the response associated with the $g$-modes is also evaluated.

Overall, the results demonstrate that space-based gravitational wave detectors such as LISA and Taiji offer a novel and complementary avenue for probing solar oscillations in near zone. A successful detection of solar $g$ modes would not only provide independent constraints on their velocity amplitudes but would also open a new observational window into the structure and dynamics of the solar core, including its rotation profile. This work highlights the potential of future space-based interferometers to contribute meaningfully to helioseismology and solar interior studies.

\section*{Acknowledgments}

I would like to express my sincere gratitude to Manan Seth for the insightful discussions during the early stages of this work. His untimely passing is deeply saddening, and I remain grateful for his valuable contributions. I also extend my thanks to B. Malavika and Siddhant Tripathy for their collaboration in the initial phases of this research. Finally, I am especially grateful to  Archana Pai for her crucial suggestions that significantly improved the final results, as well as for her thoughtful comments on the writing of this document.\\

\bibliographystyle{JHEP.bst}
\bibliography{references}

\end{document}